\RequirePackage{lineno}
\documentclass[aps,prc,twocolumn,superscriptaddress,showpacs,floatfix,longbibliography]{revtex4-2} 
\usepackage{color}
\usepackage{subfigure}
\usepackage{amsmath,epsfig}
\usepackage{amsmath}
\usepackage{graphicx}
\usepackage{epstopdf}
\usepackage{dcolumn}
\usepackage{bm}
\usepackage{multirow}
\usepackage{hhline}
\usepackage{afterpage}
\usepackage{hyperref} 
\usepackage{cleveref}
\begin{document}
\title{Identified charged hadron production in Au+Au collisions at $\sqrt{s_\mathrm{NN}}$ = 54.4~GeV with the STAR detector}
\affiliation{Academia Sinica, Nankang, 115, Taipei}
\affiliation{Abilene Christian University, Abilene, Texas   79699}
\affiliation{Alikhanov Institute for Theoretical and Experimental Physics NRC "Kurchatov Institute", Moscow 117218}
\affiliation{Argonne National Laboratory, Argonne, Illinois 60439}
\affiliation{American University in Cairo, New Cairo 11835, Egypt}
\affiliation{Ball State University, Muncie, Indiana, 47306}
\affiliation{Brookhaven National Laboratory, Upton, New York 11973}
\affiliation{University of Calabria \& INFN-Cosenza, Rende 87036, Italy}
\affiliation{University of California, Berkeley, California 94720}
\affiliation{University of California, Davis, California 95616}
\affiliation{University of California, Los Angeles, California 90095}
\affiliation{University of California, Riverside, California 92521}
\affiliation{Central China Normal University, Wuhan, Hubei 430079 }
\affiliation{University of Illinois at Chicago, Chicago, Illinois 60607}
\affiliation{Chongqing University, Chongqing, 401331}
\affiliation{Creighton University, Omaha, Nebraska 68178}
\affiliation{Czech Technical University in Prague, FNSPE, Prague 115 19, Czech Republic}
\affiliation{National Institute of Technology Durgapur, Durgapur - 713209, India}
\affiliation{ELTE E\"otv\"os Lor\'and University, Budapest, Hungary H-1117}
\affiliation{Frankfurt Institute for Advanced Studies FIAS, Frankfurt 60438, Germany}
\affiliation{Fudan University, Shanghai, 200433 }
\affiliation{Guangxi Normal University, Guilin, 541004}
\affiliation{University of Heidelberg, Heidelberg 69120, Germany }
\affiliation{University of Houston, Houston, Texas 77204}
\affiliation{Huzhou University, Huzhou, Zhejiang  313000}
\affiliation{Indian Institute of Science Education and Research (IISER), Berhampur 760010 , India}
\affiliation{Indian Institute of Science Education and Research (IISER) Tirupati, Tirupati 517507, India}
\affiliation{Indian Institute Technology, Patna, Bihar 801106, India}
\affiliation{Indiana University, Bloomington, Indiana 47408}
\affiliation{Institute of Modern Physics, Chinese Academy of Sciences, Lanzhou, Gansu 730000 }
\affiliation{University of Jammu, Jammu 180001, India}
\affiliation{Joint Institute for Nuclear Research, Dubna 141 980}
\affiliation{Kent State University, Kent, Ohio 44242}
\affiliation{University of Kentucky, Lexington, Kentucky 40506-0055}
\affiliation{Lanzhou University, Lanzhou, 730000}
\affiliation{Lawrence Berkeley National Laboratory, Berkeley, California 94720}
\affiliation{Lehigh University, Bethlehem, Pennsylvania 18015}
\affiliation{Lovely Professional University, Jalandhar - Delhi G.T. Road, Pagwara, Panjab, 144411, India}
\affiliation{Max-Planck-Institut f\"ur Physik, Munich 80805, Germany}
\affiliation{Michigan State University, East Lansing, Michigan 48824}
\affiliation{National Research Nuclear University MEPhI, Moscow 115409}
\affiliation{National Institute of Science Education and Research, HBNI, Jatni 752050, India}
\affiliation{National Cheng Kung University, Tainan 70101 }
\affiliation{The Ohio State University, Columbus, Ohio 43210}
\affiliation{Panjab University, Chandigarh 160014, India}
\affiliation{NRC "Kurchatov Institute", Institute of High Energy Physics, Protvino 142281}
\affiliation{Purdue University, West Lafayette, Indiana 47907}
\affiliation{Rice University, Houston, Texas 77251}
\affiliation{Rutgers University, Piscataway, New Jersey 08854}
\affiliation{University of Science and Technology of China, Hefei, Anhui 230026}
\affiliation{South China Normal University, Guangzhou, Guangdong 510631}
\affiliation{Sejong University, Seoul, 05006, Korea, Republic Of}
\affiliation{Shandong University, Qingdao, Shandong 266237}
\affiliation{Shanghai Institute of Applied Physics, Chinese Academy of Sciences, Shanghai 201800}
\affiliation{Southern Connecticut State University, New Haven, Connecticut 06515}
\affiliation{State University of New York, Stony Brook, New York 11794}
\affiliation{Instituto de Alta Investigaci\'on, Universidad de Tarapac\'a, Arica 1000000, Chile}
\affiliation{Temple University, Philadelphia, Pennsylvania 19122}
\affiliation{Texas A\&M University, College Station, Texas 77843}
\affiliation{Texas Southern University, Houston, Texas, 77004}
\affiliation{University of Texas, Austin, Texas 78712}
\affiliation{Tsinghua University, Beijing 100084}
\affiliation{University of Tsukuba, Tsukuba, Ibaraki 305-8571, Japan}
\affiliation{University of Chinese Academy of Sciences, Beijing, 101408}
\affiliation{Valparaiso University, Valparaiso, Indiana 46383}
\affiliation{Variable Energy Cyclotron Centre, Kolkata 700064, India}
\affiliation{Warsaw University of Technology, Warsaw 00-661, Poland}
\affiliation{Wayne State University, Detroit, Michigan 48201}
\affiliation{Wuhan University of Science and Technology, Wuhan, Hubei 430065}
\affiliation{Yale University, New Haven, Connecticut 06520}
\affiliation{University of Delhi, Delhi, India 110007}

\author{B.~E.~Aboona}\affiliation{Texas A\&M University, College Station, Texas 77843}
\author{J.~Adam}\affiliation{Czech Technical University in Prague, FNSPE, Prague 115 19, Czech Republic}
\author{G.~Agakishiev}\affiliation{Joint Institute for Nuclear Research, Dubna 141 980}
\author{I.~Aggarwal}\affiliation{Panjab University, Chandigarh 160014, India}
\author{M.~M.~Aggarwal}\affiliation{Panjab University, Chandigarh 160014, India}
\author{Z.~Ahammed}\affiliation{Variable Energy Cyclotron Centre, Kolkata 700064, India}
\author{A.~Aitbayev}\affiliation{Joint Institute for Nuclear Research, Dubna 141 980}
\author{I.~Alekseev}\affiliation{Alikhanov Institute for Theoretical and Experimental Physics NRC "Kurchatov Institute", Moscow 117218}\affiliation{National Research Nuclear University MEPhI, Moscow 115409}
\author{E.~Alpatov}\affiliation{National Research Nuclear University MEPhI, Moscow 115409}
\author{A.~K.~Alshammri}\affiliation{Kent State University, Kent, Ohio 44242}
% ...

\collaboration{STAR Collaboration}\noaffiliation




\begin{abstract}
We present results on the production of $\pi^{\pm}$, $K^{\pm}$, $p$, and $\bar{p}$ in Au+Au collisions at $\sqrt{s_\mathrm{NN}}$ = 54.4~GeV using the STAR detector at RHIC, at midrapidity ($|y| <$ 0.1). Invariant yields of these particles as a function of transverse momentum are shown. We determine bulk properties such as integrated particle yields ($dN/dy$), mean transverse momentum ($\langle p_{T} \rangle$), particle ratios, which provide insight into the particle production mechanisms. Additionally, the kinetic freezeout parameters ($T_\text{kin}$ and $\langle \beta_{T} \rangle$), which provide information about the dynamics of the system at the time of freezeout, are obtained. The Bjorken energy density ($\epsilon_{\rm{BJ}}$), which gives an estimate of the energy density in the central rapidity region of the collision zone at the formation time $\tau$, is calculated and presented as a function of multiplicity for various energies. The results are compared with those from the models such as A Multi-Phase Transport (AMPT) and Heavy Ion Jet INteraction Generator (HIJING) for further insights.
\end{abstract}

\maketitle

\section{Introduction}
Quantum chromodynamics (QCD) is a theory that describes the strong interactions occurring among quarks, mediated by gluons. Lattice QCD (LQCD) theoretically predicts a phase transition from the hadronic phase, where quarks are confined, to a deconfined phase of quarks and gluons known as quark gluon plasma (QGP)~\cite{lqcd}. The QGP is believed to have existed a few microseconds after the Big Bang, under conditions of extremely high temperature and energy density. These conditions can be achieved by colliding heavy ions at relativistic speeds. Experimental observations at high collision energies in Relativistic Heavy Ion Collider (RHIC) and Large Hadron Collider (LHC) support the existence of the QGP. These observations include strangeness enhancement~\cite{se1,se2,se3,se4,se5,se6,se7,se8}, J$/\psi$ suppression~\cite{qs1,qs2,qs3,qs4,qs5,qs6,qs7}, jet quenching~\cite{jq_star}, elliptic flow measurements~\cite{v2}, etc. 

The QCD phase diagram is a conjectured diagram usually plotted as temperature (\textit{T}) versus baryon chemical potential ($\mu_{B}$). According to LQCD calculations, at $\mu_{B}$ close to zero, the transition from QGP to a hadronic gas is a rapid crossover, whereas at larger values of $\mu_{B}$, it becomes a first-order phase transition~\cite{cce,cce2,cce3}. The point where the first-order phase transition line ends is referred to as the QCD critical point~\cite{cce4,cce5}. 

The Beam Energy Scan (BES) program at RHIC has significantly broadened the understanding of the QCD phase diagram. By changing the colliding beam energy, both \textit{T} and $\mu_{B}$ can be varied~\cite{9p2,qcd2,qcd3}. This approach presents a unique opportunity to explore several regions of the phase diagram, which include probing the phase transition regions and the possibility of finding the elusive QCD critical point. The \textit{T} and $\mu_{B}$ can be obtained from statistical thermal model analyses of the yield of produced particles~\cite{thermus,statmodel}. 

In the first phase of the BES program, the Solenoidal Tracker At RHIC (STAR) experiment collected data from Au+Au collisions at energies ranging from 7.7~GeV to 39~GeV between 2010 and 2014~\cite{BES,14p5}, and at a center-of-mass energy per nucleon-nucleon pair, $\sqrt{s_{\mathrm{NN}}}$ = 54.4~GeV in 2017. In this paper, we report the transverse momentum spectra of produced particles that provide insights into the bulk properties of the created matter, such as particle yields and ratios, mean transverse momentum, freezeout parameters, and estimates of the Bjorken energy density ($\rm{\epsilon_{BJ}}$) in Au+Au collisions at $\sqrt{s_{\mathrm{NN}}}$ = 54.4~GeV.

\section{Experimental apparatus}
The STAR detector system was constructed to investigate the behavior of strongly interacting matter at high energy density and search for the signatures of QGP. A detailed description of the detector can be found in Ref.~\cite{star_detector_system}. The time projection chamber (TPC), the heart of the STAR detector, is a large cylindrical device that sits in a solenoidal magnet operating at 0.5~Tesla~\cite{tpc}. The TPC is 4.2 m long and 4 m in diameter and covers the pseudorapidity range of $|\eta|~<~1.0$ and an azimuthal angle of 2$\pi$. It works on the principle of ionization energy loss ($dE/dx$) of charged particles passing through its gas volume. The time-of-flight (TOF) detector in STAR provides velocity information of a particle by measuring its flight time and path length~\cite{tof}. Covering full azimuthal angle and $|\eta| < 0.9$, TOF utilizes multigap resistive plate chamber (MRPC) technology to enable the identification of particles at high momentum.  	

\section{Analysis details}
The results presented in this paper are based on data taken by the STAR detector in Au+Au collisions at $\sqrt{s_{\mathrm{NN}}}$ = 54.4~GeV in 2017. The data set was taken with a minimum-bias trigger, which is defined using a coincidence of hits in the two zero degree calorimeters (ZDCs)~\cite{zdc1,zdc2} or the two vertex position detectors (VPDs)~\cite{vpds}. The primary vertex for each event was identified by determining the optimal common point from which the majority of tracks originate. To reject the background events emanating due to the interactions with the beam pipe, a cut on the event vertex radius ($V_{r} = \sqrt{V_{x}^2+V_{y}^2}$, where $V_{x}$ and $V_{y}$ are the vertex positions along $x$ and $y$ directions respectively) was applied to be less than 2~cm. To have a uniform detector acceptance, a cut of 30 cm was applied on $V_{z}$, the vertex position along the beam direction. The total number of events used for the analysis after the above-mentioned event selection cuts was approximately 500 $\times$ 10$^{6}$ .

Centrality is determined using the reference multiplicity (refmult), which is defined as the number of primary charged-particle tracks reconstructed in the TPC over the full azimuthal angle and in the pseudorapidity window $|$$\eta$$|$~$<$~0.5. This multiplicity distribution is compared and fitted with Monte Carlo Glauber model~\cite{54_npart,9p2,STAR:2008med}. Various centralities represent the fraction of the refmult. The centrality classes presented in this analysis are 0--5\% (central collisions), 5--10\%, 10--20\%, 20--30\%, 30--40\%, 40--50\%, 50--60\%, 60--70\% and 70--80\% (peripheral collisions). The mean values of the number of participating nucleons ($\langle \rm{N_{part}} \rangle$) corresponding to each centrality class were also evaluated. 

To prevent the inclusion of tracks from secondary vertices, a cut on the distance of closest approach (DCA) between each track and the event vertex of less than 3 cm was applied. Tracks must have more than 25 out of the possible 45 fit points for accurate track fitting. The number of $dE/dx$ points used to obtain $dE/dx$ values must be greater than 15. Finally, the rapidity window selected for the analysis was $|y|<$0.1 (midrapidity). 

The pions, kaons, and (anti)protons can be identified in the TPC due to their distinct mean $dE/dx$ bands when plotted as a function of rigidity $(p/q)$. The $z$ distributions defined below were used to extract the raw yields as illustrated and explained in Refs.~\cite{BES,14p5,uu,STAR:2008med}.
\begin{equation}
z_{X}=\ln\bigg( \frac{\langle dE/dx \rangle}{\langle dE/dx \rangle^B_X}\bigg),
\label{equationz}
\end{equation}
where $\textit{X}$ is the particle of interest and $\langle dE/dx\rangle^{B}_{X}$ is the theoretical energy loss predicted by the Bichsel function~\cite{bichsel,BES,STAR:2008med,uu}.
The time-of-flight information was utilized to identify the particles at relatively higher momentum $(p_{T}\sim0.4$--$2.0~\rm{GeV/\it{c}})$. The mass squared ($m^2$) distributions were obtained for all the particles for midrapidity in different $p_T$ regions which were calculated using 

\begin{equation}
m^2 = p^2\left(\frac{c^2T^2}{L^2}-1\right),
\label{m2}
\end{equation}
where, $p$, $T$, $L$, and $c$ are the momentum, time of flight of the particle, path length, and speed of light, respectively. These distributions were fitted by the predicted mass squared ones obtained using the predicted time of flight as explained in Refs.~\cite{BES,14p5,uu}. 

The raw spectra obtained using these methods were corrected for efficiency and acceptance. The TPC tracking efficiency and acceptance corrections were determined using Monte Carlo (MC) tracks simulated through the \textsc{geant3}~\cite{GEANT} model of the STAR detector, integrated into real events at the raw data level. The TOF spectra were additionally corrected for the matching efficiency using a data-driven technique~\cite{BES,14p5,uu}. The STAR track reconstruction algorithm treats all particles as pions, hence requiring a correction for energy loss for heavier particles (kaons and protons). The track $p_{T}$ for these particles has been corrected for this effect~\cite{BES,14p5,STAR:2008med,uu}. The pion spectra were corrected to account for weak decays and muon contamination. Proton spectra were corrected for secondary protons originating from the detector materials. The results for (anti)protons are inclusive, i.e., not corrected for weak decay feed down. It may be noted that the proton spectra presented here provide extended coverage toward lower $p_T$, along with finer $p_T$ (in low-momentum region) and centrality binning compared to those reported earlier~\cite{54p4_p_spectra}. The analysis methods and all correction procedures used for the particles in this study are the same as those described in Refs.~\cite{BES,14p5}.

\begin{figure*}[htbp]
\begin{center}
\includegraphics[scale=0.807]{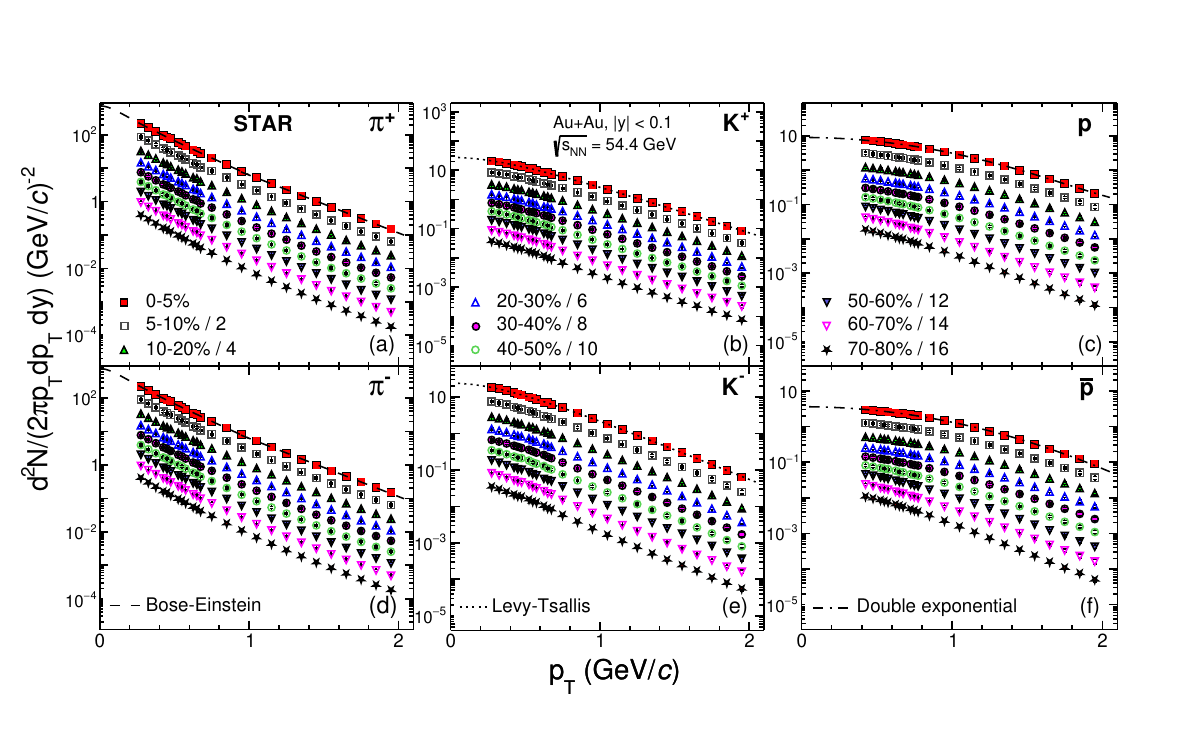}
\vspace{-0.5cm}
\caption{The transverse momentum spectra for (a) $\pi^{+}$, (b) $K^{+}$, (c) $p$, (d) $\pi^{-}$, (e) $K^{-}$ and (f) $\bar{p}$ at midrapidity ($ |y| < 0.1$) in Au+Au collisions at $\sqrt{s_\mathrm{NN}}$ = 54.4~GeV for nine centrality classes. The spectra for all the centralities other than 0--5\% are scaled for clarity. The curves represent the Bose-Einstein functional fit to pions, Levy-Tsallis fit to kaons, and double-exponential fit to protons and antiprotons for 0--5\% centrality. The statistical and systematic uncertainties are added in quadrature where the latter dominates.}
\label{spectrafig}
\end{center}
\end{figure*}

\begin{figure*}[htbp]
\begin{center}
\includegraphics[scale=0.8]{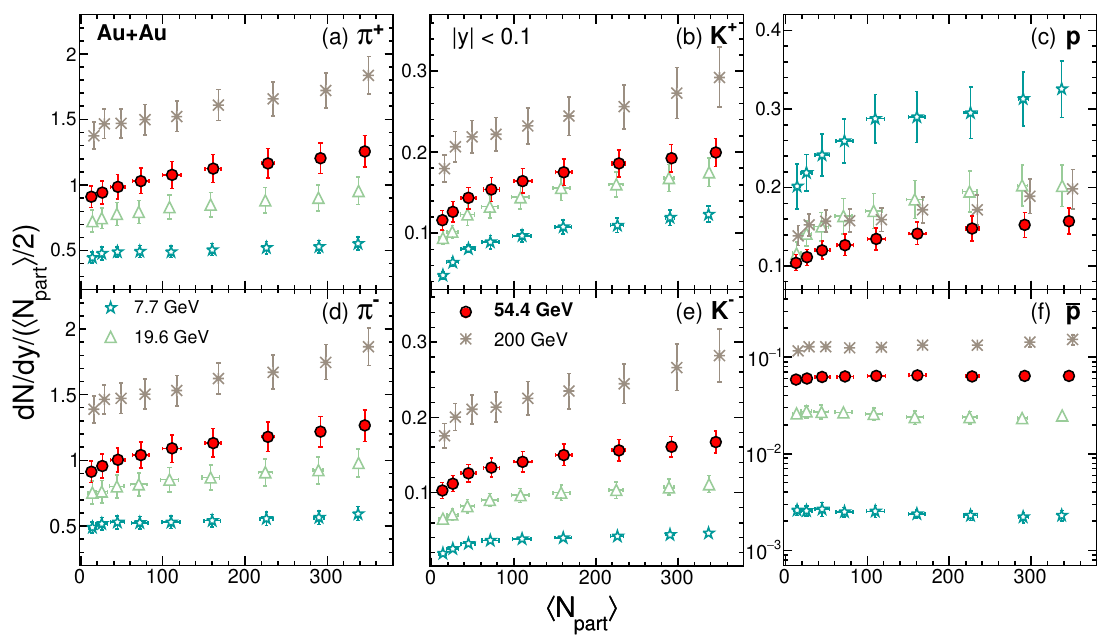}
\vspace{-0.5cm}
\caption{The $\langle \rm{N_{part}}\rangle$ dependence of the normalized integrated particle yield ($dN/dy/(\langle \rm{N_{part}}\rangle/2$) for (a) $\pi^{+}$, (b) $K^{+}$, (c) $p$, (d) $\pi^{-}$, (e) $K^{-}$, (f) $\bar{p}$ at midrapidity ($ |y| < 0.1$) in Au+Au collisions at $\sqrt{s_\mathrm{NN}}$ = 54.4~GeV. The results are compared with the published results at other STAR energies~\cite{BES,STAR:2008med}. The statistical and systematic uncertainties are added in quadrature where the latter dominates. $\rm \langle N_{part}\rangle$ uncertainties are not combined in quadrature for clarity of plots.}
\label{dndy_cent}
\end{center}
\end{figure*}

The systematic uncertainties on the spectra were estimated by varying the event and the track selection cuts from their default values. The variations were done for the vertex selection cuts ($V_z$), DCA, number of fit points, number of $dE/dx$ points, and the PID cut~\cite{BES}. The uncertainty from each cut was calculated as the difference between the default value and that obtained with the varied cut. The systematic uncertainty in the proton background was calculated using the similar methods as in Ref.~\cite{BES} and it contributed approximately  5--6\%. The uncertainty due to the pion background is negligible and that due to the track reconstruction efficiency and acceptance is estimated to be 5\%~\cite{BES,14p5,STAR:2008med}. The yields in the unmeasured $p_{T}$ regions were obtained by extrapolating the functional fits to the spectra. For pions, a Bose-Einstein function was the default function, while a $p_{T}$ exponential was used as the varied function. For kaons, the Levy-Tsallis~\cite{levy} function was the default while the Boltzmann function or $m_{T}$ exponential were used as varying functions. For (anti)protons, the double-exponential function was the default function and an $m_T$ exponential function was used as the varied function. These functions were selected as they provide the best description of the corresponding spectra and also by taking guidance from the previous published papers~\cite{BES,14p5,STAR:2008med}. The percentage of extrapolation in the particle yields was 36--40\% for pions, 15--20\% for kaons, and 16--29\% for protons and antiprotons across different centralities. The contributions from the different sources were added in quadrature and the resultant total systematic uncertainties on particle yields are summarized in Table \ref{table:1}. The systematic uncertainties on particle ratios were estimated by propagating those from particle yields, where the correlated uncertainty i.e. due to tracking efficiency was excluded. In addition, the extrapolation uncertainties cancel in antiparticle to particle ratios. The systematic uncertainties on $\langle p_{T} \rangle$ were dominated by those from extrapolation as discussed above. The total systematic uncertainties on $\langle p_{T} \rangle$ for pions were 5--6\% and for kaons and (anti)protons were 6--7\%. The systematic uncertainties might have some correlation among centralities. The uncertainties in the $\rm{\epsilon_{BJ}}$ $\times$ $\tau$ and transverse overlap area of the two colliding nuclei ($S_\perp$) were calculated by error propagation [see eqs.~\ref{eq1} - \ref{eq3}]~\cite{STAR:2008med}. 

\begin{table}
\begin{center}
\begin{tabular}{| c | c | c | c|}
	\hline
	Sources & $\pi^{\pm}$ & $K^{\pm}$ & $p$($\bar{p}$)\\
	\hline 
	Cuts & 4\% & 3\% & 6\%\\
	\hline
	Tracking efficiency & 5\% & 5\% & 5\%\\
	\hline
	Extrapolation & 6--7\% & 6--8\% & 4--8\%\\
	\hline
	Total & 9--10\% & 8--10\% & 9--11\%\\
	\hline
\end{tabular}
\caption{Sources of systematic uncertainties for pions, kaons and (anti)protons yields in  Au+Au collisions at  $\sqrt{s_\mathrm{NN}}$ = 54.4~GeV.} 
\label{table:1}
\end{center}
\end{table}

\section{Results and discussion}
The transverse momentum spectra of $\pi^{\pm}$, $K^{\pm}$, $p$ and $\bar{p}$ in Au+Au collisions at $\sqrt{s_\mathrm{NN}}$ = 54.4~GeV at midrapidity ($|y| < 0.1$) are shown in Fig.~\ref{spectrafig}. The invariant yield decreases with increasing $p_{T}$ and towards peripheral collisions. Moreover, the flattening of the spectra increases with mass and towards central collisions, a phenomenon attributed to the radial flow~\cite{BES,STAR:2008med,14p5}. These spectra are fitted with functions as shown in figure to obtain the integrated particle yields $(dN/dy)$ and mean transverse momentum($\langle p_{T} \rangle$). The dependence of $dN/dy$, $\langle p_{T} \rangle$, particle ratios etc. on centrality and energy give more information on the particle production mechanisms and bulk properties of the system. 

The integrated particle yield obtained for each particle is normalized by half the number of participating nucleons in the collision centrality, facilitating a comparative study across different centralities. Figure~\ref{dndy_cent} shows the $dN/dy/(\langle \rm{N_{part}}\rangle/2)$ for all the particles [Fig. 2(a) $\pi^{+}$, 2(b) $K^{+}$, 2(c) $p$, 2(d) $\pi^{-}$, 2(e) $K^{-}$, 2(f) $\bar{p}$] as a function of $\langle \rm{N_{part}}\rangle$. The results for Au+Au collisions at $\sqrt{s_\mathrm{NN}}$ = 54.4~GeV at midrapidity ($|y| < 0.1$) are shown in comparison to other STAR energies~\cite{BES,STAR:2008med}. The normalized yields for $\pi^{+}$, $\pi^{-}$, $K^{+}$, $K^{-}$ and $p$ show a clear centrality dependence, with the yield increasing towards central collisions, indicating the contributions from both soft and hard processes involving nucleon-nucleon binary collisions. For antiprotons the centrality dependence is weak, indicating the increase of baryon-antibaryon annihilation effects towards central collisions~\cite{14p5}. The yields of pions, kaons and antiprotons increase with increasing energy. The proton yield decreases with increasing energy from $\sqrt{s_\mathrm{NN}}$ = 7.7~GeV till 54.4~GeV and then increases at 200~GeV. The low proton yield at $\sqrt{s_\mathrm{NN}} = 54.4$~GeV can be attributed to the interplay between baryon stopping and pair-production mechanisms~\cite{BES}.

\begin{figure*}[htbp]
\begin{center}
\includegraphics[scale=0.8]{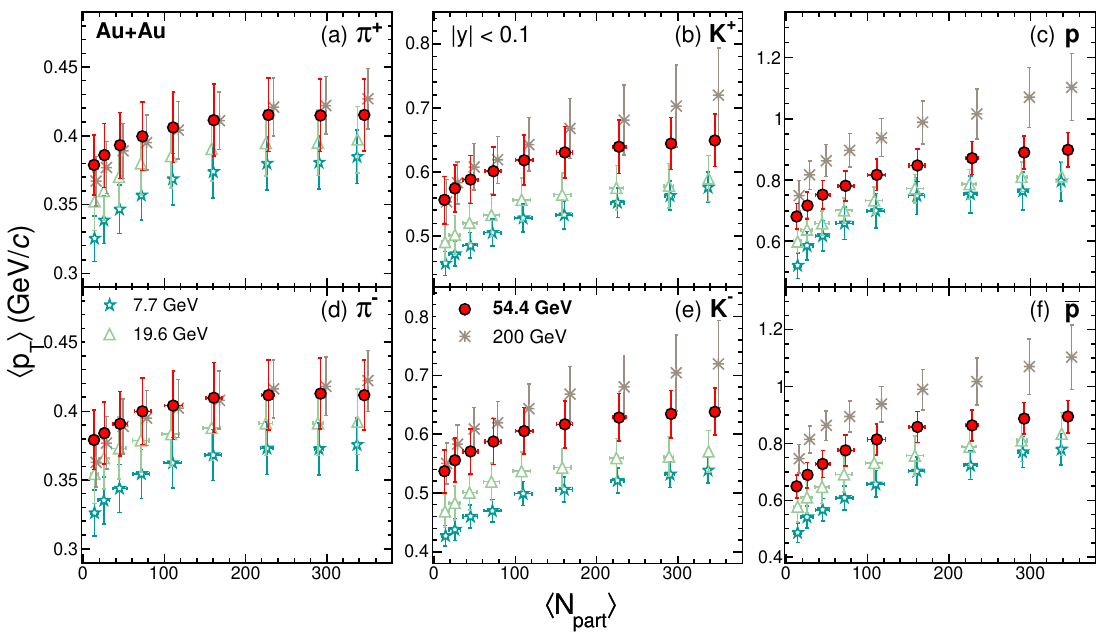}
\vspace{-0.5cm}
\caption{The $\langle \rm{N_{part}}\rangle$ dependence of mean transverse momentum ($\langle p_{T} \rangle$) of (a) $\pi^{+}$, (b) $K^{+}$, (c) $p$, (d) $\pi^{-}$, (e) $K^{-}$, (f) $\bar{p}$ at midrapidity ($ |y| < 0.1$) in Au+Au collisions at $\sqrt{s_\mathrm{NN}}$ = 54.4~GeV. The results are compared with the published results at other STAR energies~\cite{BES,STAR:2008med}. The statistical and systematic uncertainties are added in quadrature where the latter dominates.}
\label{meanpt_a}
\end{center}
\end{figure*}

\begin{figure*}[htbp]
\begin{center}
\includegraphics[scale=0.8]{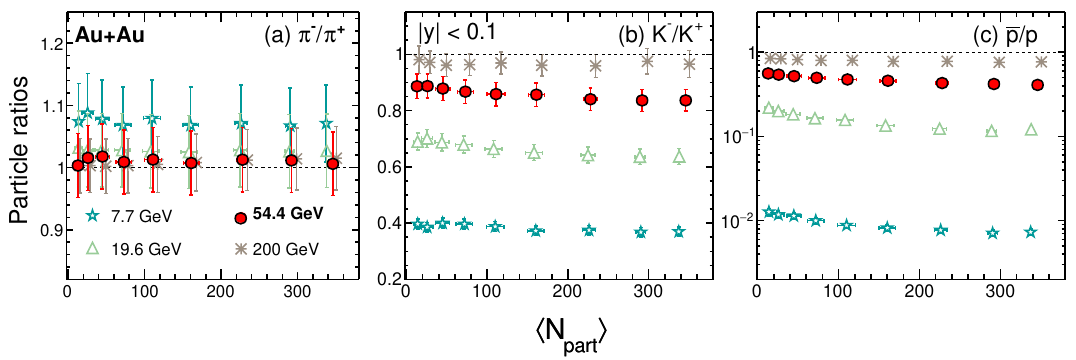}
\vspace{-0.5cm}
\caption{The $\langle \rm{N_{part}}\rangle$ dependence of the antiparticle to particle ratios (a) $\pi^{-}$/$\pi^{+}$, (b) $K^{-}$/$K^{+}$ and (c) $\bar{p}$/$p$ at midrapidity ($|y| <$ 0.1) in Au+Au collisions at $\sqrt{s_\mathrm{NN}}$ = 54.4~GeV. The results are compared with the published results at other STAR energies~\cite{BES,STAR:2008med}. The statistical and systematic uncertainties are added in quadrature where the latter dominates.}
\label{dndyapcpc}
\end{center}
\end{figure*}

\begin{figure*}[htbp]
\begin{center}
\includegraphics[scale=0.6]{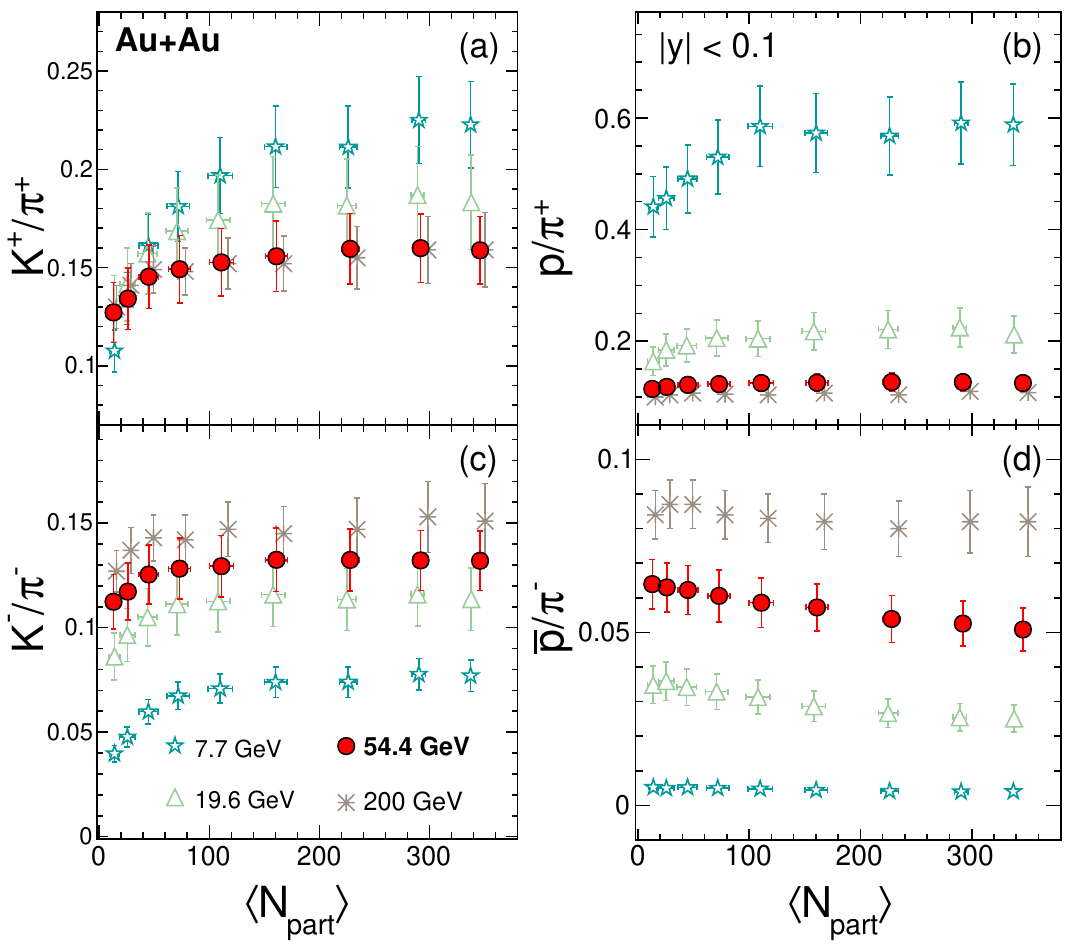}
\vspace{-0.5cm}
\caption{The $\langle \rm{N_{part}}\rangle$ dependence of mixed ratios (a) $K^{+}$/$\pi^{+}$, (b) ${p}$/$\pi^{+}$, (c) $K^{-}$/$\pi^{-}$ and (d) $\bar{p}$/$\pi^{+}$ at midrapidity ($|y| <$ 0.1) in Au+Au collisions at $\sqrt{s_\mathrm{NN}}$ = 54.4~GeV. The results are compared with the published results at other STAR energies~\cite{BES,STAR:2008med}. The statistical and systematic uncertainties are added in quadrature where the latter dominates.}
\label{mixedra}
\end{center}
\end{figure*}

Figure~\ref{meanpt_a} shows the dependence of the mean transverse momentum on $\langle \rm{N_{part}}\rangle$ for $\pi^{+}$, $\pi^{-}$, $K^{+}$, $K^{-}$, $p$, and $\bar{p}$ in Au+Au collisions at $\sqrt{s_\mathrm{NN}}$ = 54.4~GeV. These results are compared with results at other STAR energies~\cite{BES, STAR:2008med}. It is observed that the mean transverse momentum increases towards more central collisions for all the particles and this trend is consistent across all energies. In addition, the $\langle p_{T} \rangle$ shows an increase with increasing mass, following the order, $\pi^{+}(\pi^{-}) < K^{+}(K^{-}) < p(\bar{p})$. These features suggest the existence of radial flow in these collisions.

Figure~\ref{dndyapcpc} shows the antiparticle to particle ratios [Fig. 4(a) $\pi^{-}$/$\pi^{+}$, 4(b) $K^{-}$/$K^{+}$, and 4(c) $\bar{p}$/$p$] as a function of $\langle \rm{N_{part}}\rangle$. The results obtained in Au+Au collisions at $\sqrt{s_\mathrm{NN}}$ = 54.4~GeV at midrapidity ($|y| < 0.1$) is shown in comparison to other STAR energies~\cite{BES,STAR:2008med}. The $\pi^{-}$/$\pi^{+}$ ratio is close to unity for all centralities in Au+Au collisions at $\sqrt{s_\mathrm{NN}}$ = 54.4~GeV. For other energies presented in figure the ratio is close to 1 except $\sqrt{s_\mathrm{NN}}$ = 7.7~GeV which is greater than 1 for all centralities. The $K^{-}$/$K^{+}$ ratio is almost flat with centrality but increases with increasing energy. The $\bar{p}/p$ ratio increases slightly from central to peripheral collisions which reflects high baryon stopping at midrapidity and/or baryon-antibaryon annihilation in more central collisions as compared to the peripheral collisions. The ratio increases with increasing collision energy.

Figure~\ref{mixedra} shows the $\langle \rm{N_{part}}\rangle$ dependence of ratios [Fig. 5(a) $K^{+}$/$\pi^{+}$, 5(b) $p$/$\pi^{+}$, 5(c) $K^{-}$/$\pi^{-}$, and 5(d) $\bar{p}$/$\pi^{-}$]. The results are compared with the published results~\cite{BES,STAR:2008med}. The $K^{\pm}$/$\pi^{\pm}$ ratios increase from peripheral to mid central and then become constant towards central collisions. The $K^{-}/\pi^{-}$ ratio increases with increasing collision energy. For $K^{+}/\pi^{+}$, the increase from peripheral to central collisions is steeper at lower energies, with the highest value observed in central collisions at $\sqrt{s_{NN}}$ = 7.7~GeV. The $p$/$\pi^{+}$ ratio increases from peripheral to central collisions for lower energies while no significant variation is observed at energies of 54.4~GeV and above. The ratio is the highest at 7.7~GeV and decreases with increasing energy, which is a consequence of large baryon stopping at lower energies. The $\bar{p}$/$\pi^{-}$ slightly increases from central to peripheral collisions, and increases with increasing energy.

\begin{figure*}[htbp]
\begin{center}
\includegraphics[scale=0.9]{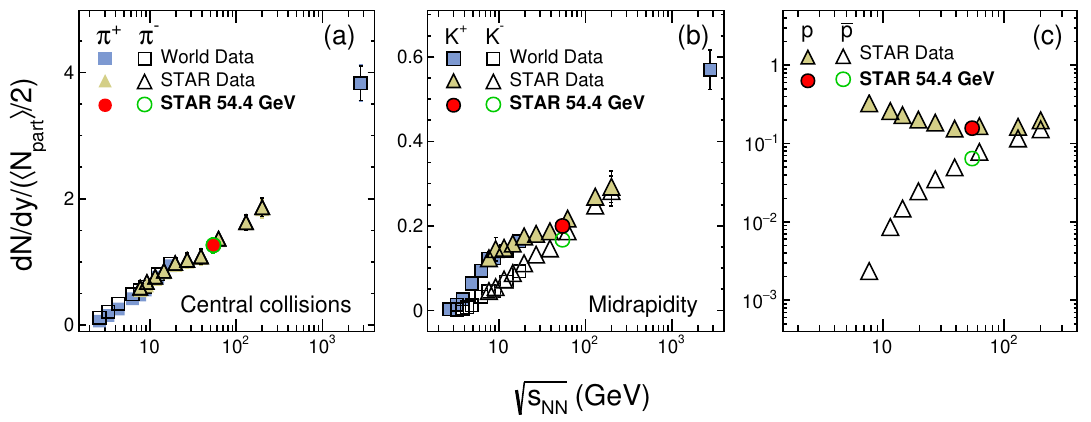}
\vspace{-0.5cm}
\caption{The energy dependence of the normalized integrated particle yield ($dN/dy/(\langle \rm{N_{part}}\rangle/2)$) at midrapidity for (a) $\pi^{\pm}$, (b) $K^{\pm}$ and (c) $p$ and $\bar{p}$. Results  in 0--5\% Au+Au collisions at $\sqrt{s_\mathrm{NN}}$ = 54.4~GeV are compared with the published results of most central collisions from AGS~\cite{AGS1,AGS2,AGS3,AGS4,AGS5,AGS6,AGS7,AGS8}, SPS~\cite{SPS1,SPS2,SPS3,SPS4}, RHIC~\cite{BES,STAR:2008med,phe,9p2}, and LHC~\cite{alice2p76}. Uncertainties shown are the quadrature sum of statistical and systematic uncertainties where the latter dominates.}
\label{dndysnn}
\end{center}

\end{figure*}
\begin{figure*}[htbp]
\begin{center}
\includegraphics[scale=0.9]{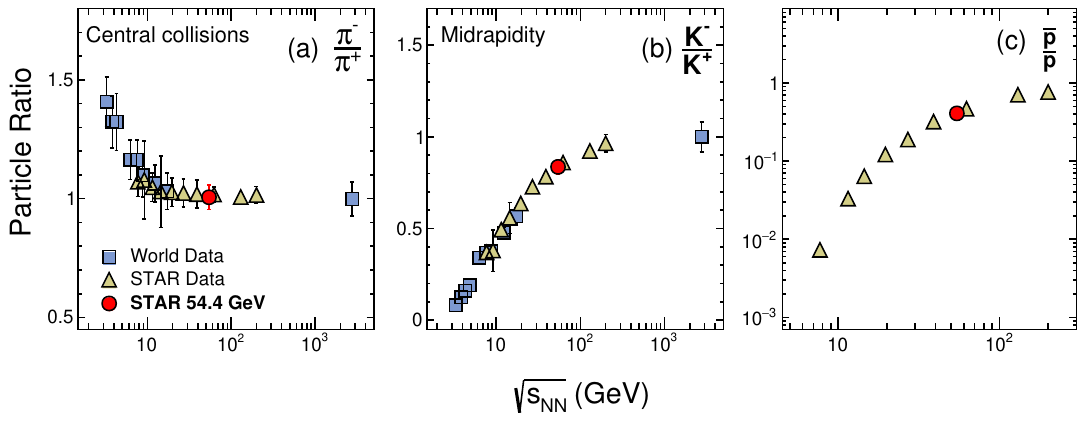}
\vspace{-0.5cm}
\caption{The energy dependence of the antiparticle to particle ratios for $(a)\pi^{-}$/$\pi^{+}$, (b) $K^{-}$/$K^{+}$ and (c) $\bar{p}$/$p$. Results in 0--5\% Au+Au collisions $\sqrt{s_\mathrm{NN}}$ = 54.4~GeV are compared with the published results of most central collisions from AGS~\cite{AGS1,AGS2,AGS3,AGS4,AGS5,AGS6,AGS7,AGS8}, SPS~\cite{SPS1,SPS2,SPS3,SPS4}, RHIC~\cite{BES,STAR:2008med,9p2}, and LHC~\cite{alice2p76}. Uncertainties shown are the quadrature sum of statistical and systematic uncertainties where the latter dominates.}
\label{dndysnnratio}
\end{center}
\end{figure*}

Figure~\ref{dndysnn} shows the energy dependence of the normalized integrated particle yield for $\pi^{\pm}$, $K^{\pm}$, $p$, and $\bar{p}$ at midrapidity $(|y|<0.1)$ as a function of collision energy. The results for the analysis of Au+Au most central collisions at $\sqrt{s_\mathrm{NN}}$ = 54.4~GeV are in agreement with the trend observed for the central collisions at other STAR energies~\cite{BES,STAR:2008med,9p2} and the world data~\cite{AGS1,AGS2,AGS3,AGS4,AGS5,AGS6,AGS7,AGS8,SPS1,SPS2,SPS3,SPS4,alice2p76}. The energy dependence of the normalized integrated particle yield for $\pi^{\pm}$ increases with energy. However there seems to be a slight change of slope around 19.6~GeV, suggesting a difference in the particle production mechanism around this energy~\cite{BES}. A significant difference between the yield of $K^{+}$ and $K^{-}$ can be observed at low energies. This might be due to the dominance of associated production at low energies, resulting in an increase in the yield of $K^{+}$ as compared to $K^{-}$. Associated production involves reactions such as $NN \to KYN$ and $\pi N \to KY$, where $N$ represents a nucleon and $Y$ denotes a hyperon~\cite{BES}. The effect of associated production decreases with increasing energy and the pair production mechanism takes over as the dominating mechanism for particle production at higher energies. The energy dependence of proton yields decreases with increasing energy which eventually saturates at higher energies. The large yield of protons at lower energies is a result high baryon stopping at lower energies~\cite{BES,9p2,barstp}, the effect of which decreases with increasing energy. The energy dependence of $\bar{p}$ yield, however, shows a steady increase with the increasing energy.

\begin{figure}[htbp]
\includegraphics[scale=0.46]{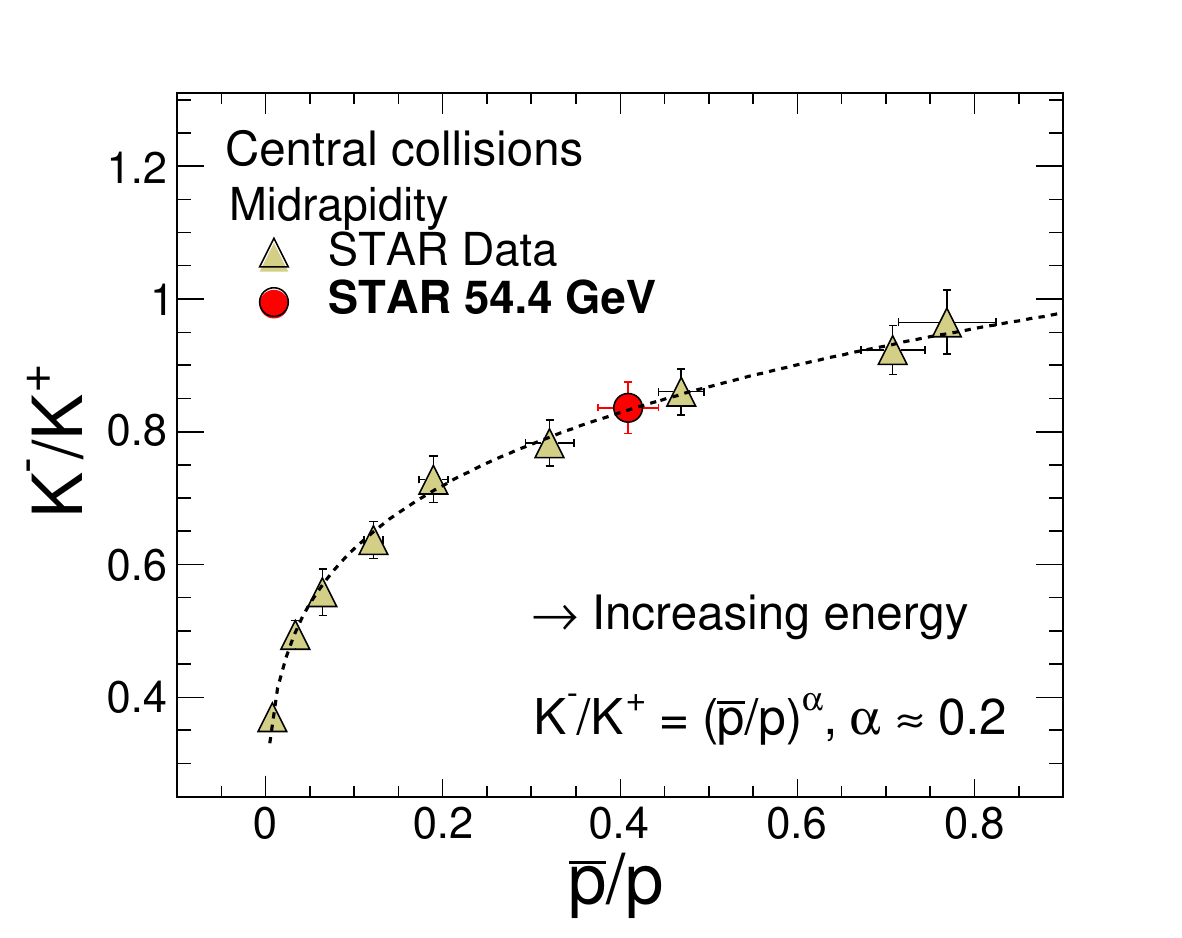}
\vspace{-0.5cm}
\caption{The dependence of the $K^{-}$/$K^{+}$ as a function of $\bar{p}$/$p$. The plot is fitted with a power-law function. Results in 0--5\% Au+Au collisions $\sqrt{s_\mathrm{NN}}$ = 54.4~GeV are compared with the published results for the most central collisions from STAR~\cite{BES,STAR:2008med}. Uncertainties shown are the quadrature sum of statistical and systematic uncertainties where the latter dominates.}
\label{extrapa}
\end{figure}

\begin{figure}[htbp]
\includegraphics[scale=0.47]{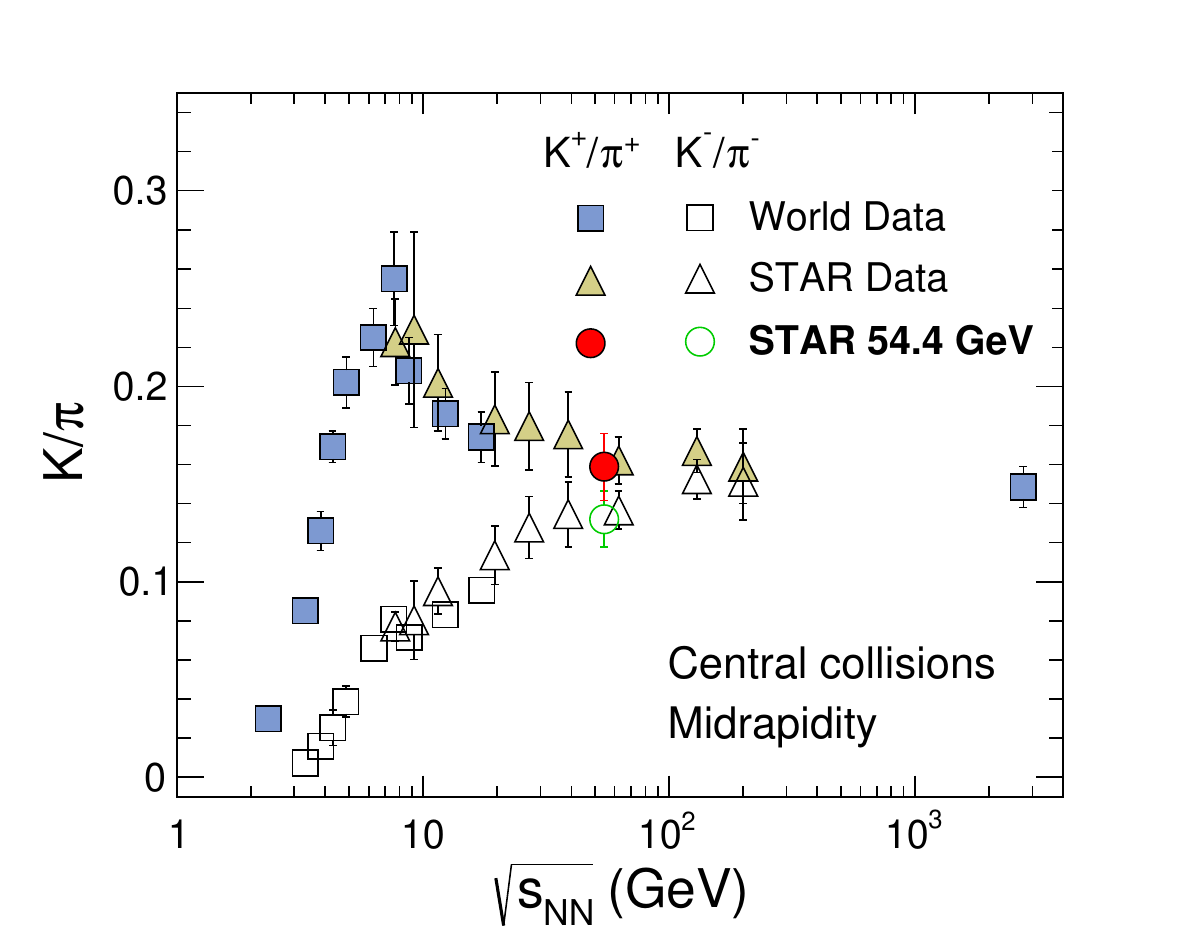}
\vspace{-0.5cm}
\caption{The energy dependence of the $K$/$\pi$ ratio is shown. Results in 0--5\% Au+Au collisions $\sqrt{s_\mathrm{NN}}$ = 54.4~GeV are compared with the published results of the most central collisions from AGS~\cite{AGS1,AGS2,AGS3,AGS4,AGS5,AGS6,AGS7,AGS8}, SPS~\cite{SPS1,SPS2,SPS3,SPS4}, RHIC~\cite{BES,STAR:2008med,9p2}, and LHC~\cite{alice2p76}. Uncertainties shown are the quadrature sum of statistical and systematic uncertainties where the latter dominates.}
\label{extrapb}
\end{figure}

\begin{figure*}[htbp]
\begin{center}
\includegraphics[scale=0.9]{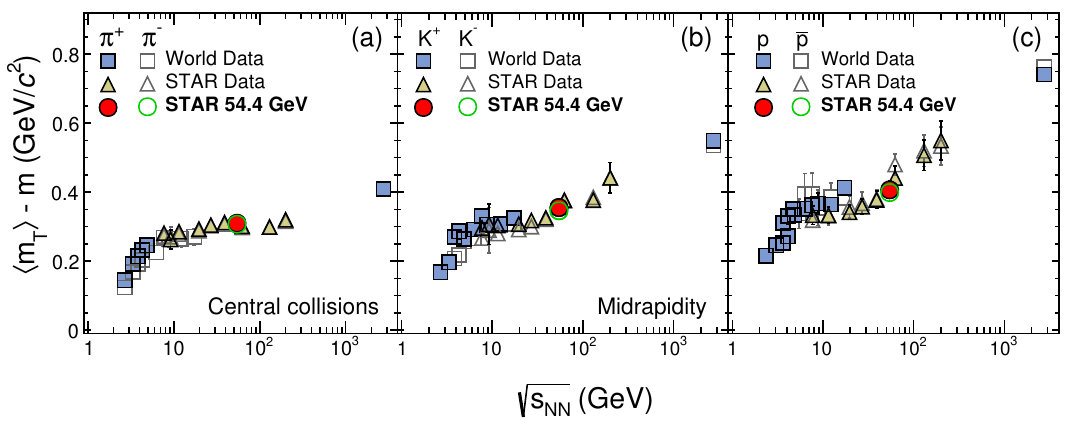}
\vspace{-0.5cm}
\caption{The energy dependence of the $\langle m_{T} \rangle - m $ in (a) $\pi^{\pm}$, (b) $K^{\pm}$ and (c) $p$ and $\bar{p}$ is shown. Results in 0--5\% Au+Au collisions $\sqrt{s_\mathrm{NN}}$ = 54.4~GeV are compared with the published results for most central collisions from AGS~\cite{AGS1,AGS2,AGS3,AGS4,AGS5,AGS6,AGS7,AGS8}, SPS~\cite{SPS1,SPS2,SPS3,SPS4}, RHIC~\cite{BES,STAR:2008med,9p2}, and LHC~\cite{alice2p76}. Uncertainties shown are the quadrature sum of statistical and systematic uncertainties where the latter dominates.}
\label{meanmtall}
\end{center}
\end{figure*}

The antiparticle to particle ratios as a function of collision energy are shown in Fig.~\ref{dndysnnratio}. The results from Au+Au most central collisions (0--5\%) at $\sqrt{s_\mathrm{NN}}$ = 54.4~GeV are in agreement with the trend observed for STAR energies and the other world data. For the most central Au+Au collisions at $\sqrt{s_\mathrm{NN}}$ = 54.4~GeV, the antiparticle to particle ratio for pions is approximately 1; for kaons it is close to 0.84 and for protons it is approximately 0.40. The $\pi^{-}$/$\pi^{+}$ ratio is greater than unity at lower energies. This might be due to the isospin and contributions from resonance decays ($\Delta$ baryons) at lower energies~\cite{BES}. The ratio gradually approaches unity at higher energies. The antiparticle to particle ratio for kaons (i.e. $K^{-}$/$K^{+}$) increases with increasing energy. The low value of the ratio at lower energies might be due to the dominance of associated production as a mechanism for particle production which results in the increased yield of $K^{+}$. However, at higher energies, the pair production dominates. The $\bar{p}/p$ ratio is lower at low energies which reflects large proton yields as compared to antiprotons due to high baryon density at midrapidity at lower energies. 

Figure~\ref{extrapa} shows the variation of $K^{-}$/$K^{+}$ as a function of $\bar{p}$/$p$ for most central collisions at STAR energies~\cite{BES,14p5,STAR:2008med}. In a hadron gas, the relationship between the strange and baryon chemical potentials is temperature dependent~\cite{ratios_kp}. This plot estimates how the net strange chemical potential (related to kaon production) is connected to the net baryon density (indicated by the antiproton-proton ratio). The correlation between $K^{-}$/$K^{+}$ and $\bar{p}$/$p$ ratio follows a power-law behavior, $K^{-}$/$K^{+}$ = ($\bar{p}$/$p)^{\alpha}$ with the value of the parameter $\alpha$ approximately 0.2. Both the ratios approach unity at higher energies. Figure~\ref{extrapb} shows the energy dependence of $K$/$\pi$ ratio. The results are compared with the other STAR energies~\cite{BES,STAR:2008med} and the world data~\cite{AGS1,AGS2,AGS3,AGS4,AGS5,AGS6,AGS7,AGS8,SPS1,SPS2,SPS3,SPS4,alice2p76}. This ratio reflects the strangeness content relative to entropy in heavy-ion collisions. The $K$/$\pi$ ratio at 54.4~GeV follows the trend with the published data. The ratio $K^{-}$/$\pi^{-}$ shows a steady increase at lower energies, becoming almost constant at higher energies. The $K^{+}$/$\pi^{+}$ ratio does not show a similar trend, but rather has a peak like structure commonly known as the horn.  The peak in $K^{+}$/$\pi^{+}$ ratio around 7.7~GeV might be related to the maximum net baryon density predicted around this energy~\cite{cley}. 

Figure~\ref{meanmtall} shows the energy dependence of $\langle m_{T}\rangle-m$ for Fig. 10(a) $\pi^{\pm}$, 10(b) $K^{\pm}$, 10(c) $p$ and $\bar{p}$. Here, $m_T$ represents the transverse mass defined as $\sqrt{p_T^2+m^2}$. $\langle m_{T}\rangle-m$ is said to be the approximate representation of the temperature of the system, and $dN/dy$ $\propto$ ln($\sqrt{s_\mathrm{NN}}$) might represent its entropy~\cite{BES}. $\langle m_{T}\rangle - m$ increases with increasing $\sqrt{s_\mathrm{NN}}$ at lower energies, and then becomes constant around the BES energies (7.7 -- 39~GeV) and then rises further. This could reflect the formation of mixed phase of QGP and hadrons as suggested by Van Hove~\cite{vanhove}. However, at $\sqrt{s_\mathrm{NN}} = $54.4~GeV, $\langle m_{T}\rangle - m$ is above this constant trend. The slight increase in the observable at 54.4~GeV might indicate a deconfined phase. However, $\langle m_{T}\rangle - m$ can also be influenced by several other effects, which must be understood for a proper interpretation of the data~\cite{BMmT}.

The chemical freezeout condition refers to a point during the evolution of the heavy-ion collisions where the particle production process ceases. The kinetic freezeout, also known as thermal freezeout is the stage at which the system cools down to a point where the elastic interactions among the particles become negligible. The various freezeout parameters can be extracted from the abundances of various particles and their transverse momentum distributions. The determination of chemical freezeout parameters is beyond the scope of this paper due to the absence of sufficient data on the yield of strange hadron particles at the energy level under investigation. However, it may be considered in future studies.

\begin{figure}[htbp]
\begin{center}
\includegraphics[scale=0.42]{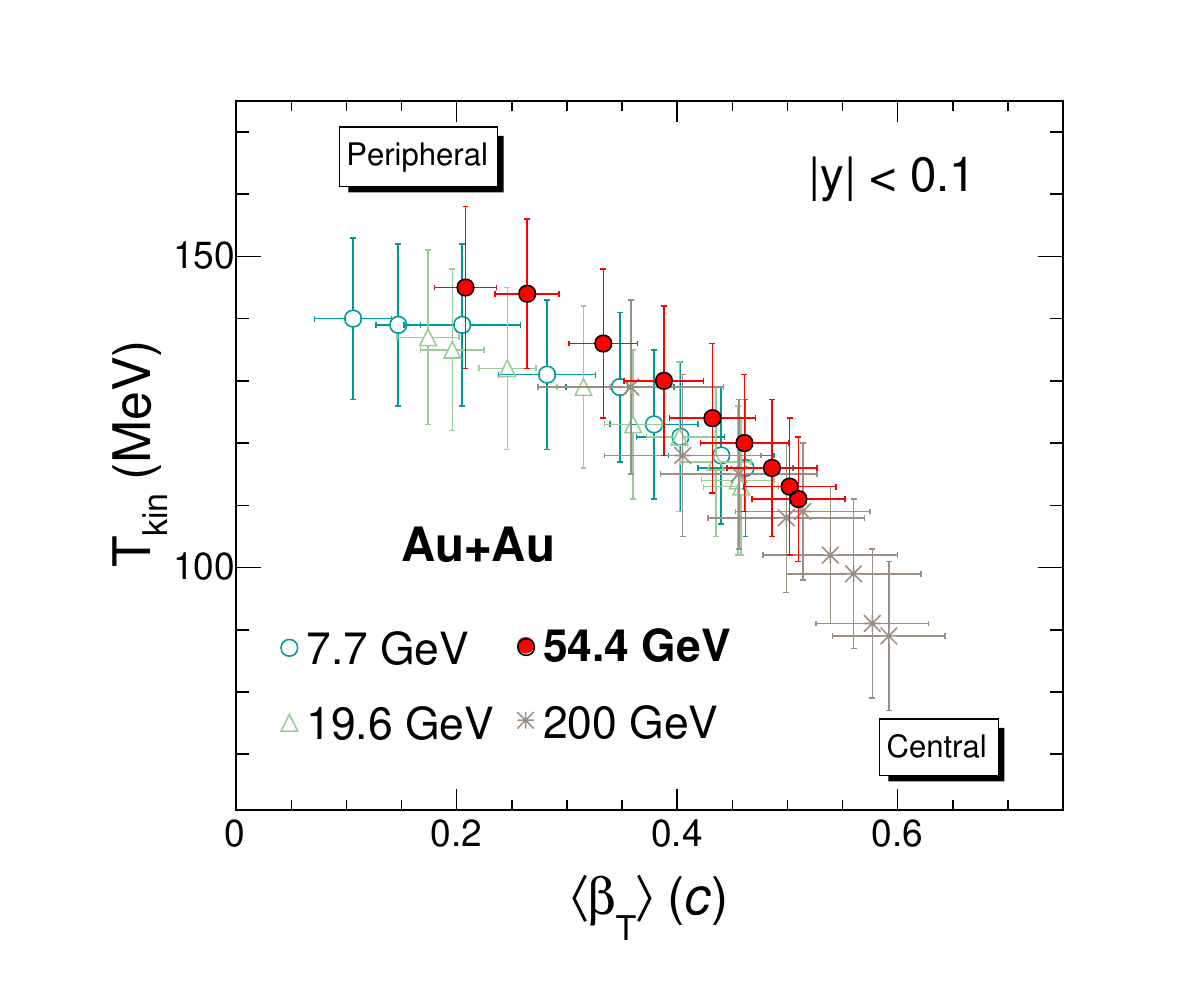}
\vspace{-0.5cm}
\caption{The dependence of the freezeout parameters $T_\text{kin}$ with $\langle \beta_{T} \rangle$. Results in Au+Au collisions $\sqrt{s_\mathrm{NN}}$ = 54.4~GeV are compared with the published results for STAR energies~\cite{BES,STAR:2008med}. Uncertainties shown are the systematic uncertainties.}
\label{tkinbeta}
\end{center}
\end{figure}

\begin{figure}[htbp]
\includegraphics[scale=0.45]{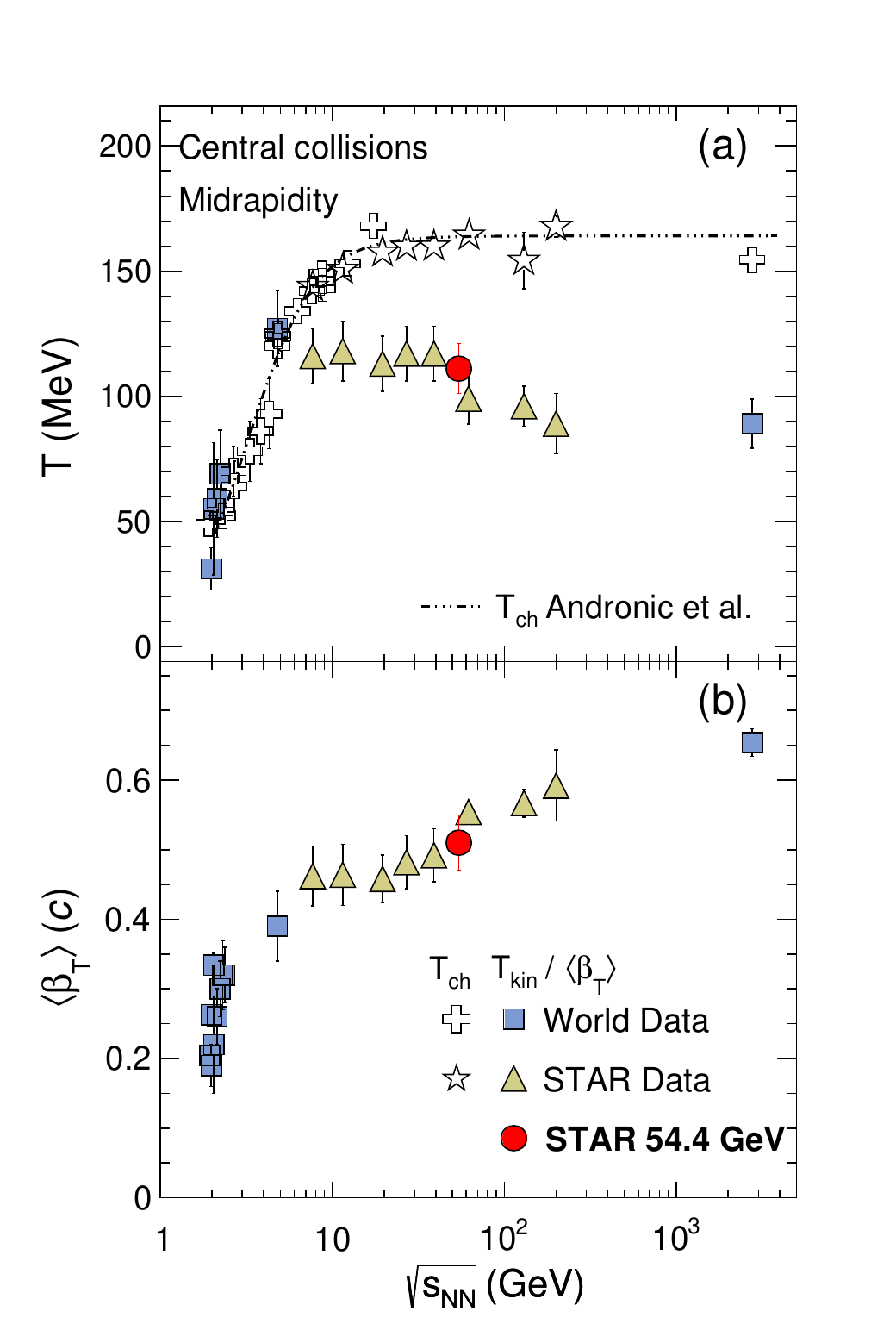}
\vspace{-0.5cm}
\caption{The energy dependence of (a) $T_\text{kin}$ and $T_\text{ch}$, and (b) $\langle \beta_{T} \rangle$. The curve represents the theoretical prediction~\cite{thcur}. The results obtained for midrapidity in Au+Au collisions at $\sqrt{s_\mathrm{NN}}$ = 54.4~GeV are compared with the results of the most central collisions at other energies~\cite{AGS1,AGS2,AGS3,AGS4,AGS5,AGS6,AGS7,AGS8,SPS1,SPS2,SPS3,SPS4,BES,STAR:2008med,alice2p76}. Uncertainties shown are systematic uncertainties.}
\label{tkinbetasnn}
\end{figure}

To obtain the kinetic freezeout parameters, the hydrodynamics-motivated blast-wave model is used~\cite{STAR:2008med,bw2,BES}. The transverse momentum spectra of $\pi^{\pm}$, $K^{\pm}$, $p$, and $\bar{p}$ are fitted simultaneously to extract the kinetic freezeout temperature ($T_\text{kin}$) and average transverse radial flow velocity ($\langle \beta_{T} \rangle$). The systematic uncertainties described in Sec. III are included in these spectra for the blast wave fits. The functional form of the blast-wave and the fit ranges used for each particle are the same as used in Ref.~\cite{BES}. Figure~\ref{tkinbeta} shows the variation of $T_\text{kin}$ with $\langle \beta_{T} \rangle$ and the results are compared with other STAR energies~\cite{BES,STAR:2008med}. These parameters show an anticorrelation with each other. It can be observed that $T_\text{kin}$ increases from central to peripheral collisions which suggests a longer-lived fireball in central collisions. However, $\langle \beta_{T} \rangle$ decreases from central to peripheral collisions, indicating a rapid expansion in central collisions. The variation of the freezeout parameters as a function of energy is shown in Fig.~\ref{tkinbetasnn}. The results obtained for the kinetic freezeout parameters at midrapidity in Au+Au collisions at $\sqrt{s_\mathrm{NN}} = 54.4$~GeV are compared with the results of the most central collisions at other energies~\cite{AGS1,AGS2,AGS3,AGS4,AGS5,AGS6,AGS7,AGS8,SPS1,SPS2,SPS3,SPS4,BES,STAR:2008med,alice2p76}. The dependence of the obtained kinetic freezeout parameters on energy follow the world data trend. It can be observed that at low energies the kinetic and chemical freezeout temperatures are comparable, while at higher energies a significant gap can be observed which increases with increasing energy.  This might be due to the increasing hadronic interactions between chemical and kinetic freezeout at higher energies~\cite{tkin_low}. The average transverse radial flow velocity ($\langle \beta_{T} \rangle$) increases steeply at lower energies, and steadily beyond low RHIC energies up to LHC energies. 

\begin{figure*}[htbp]
\begin{center}
\includegraphics[scale=0.7]{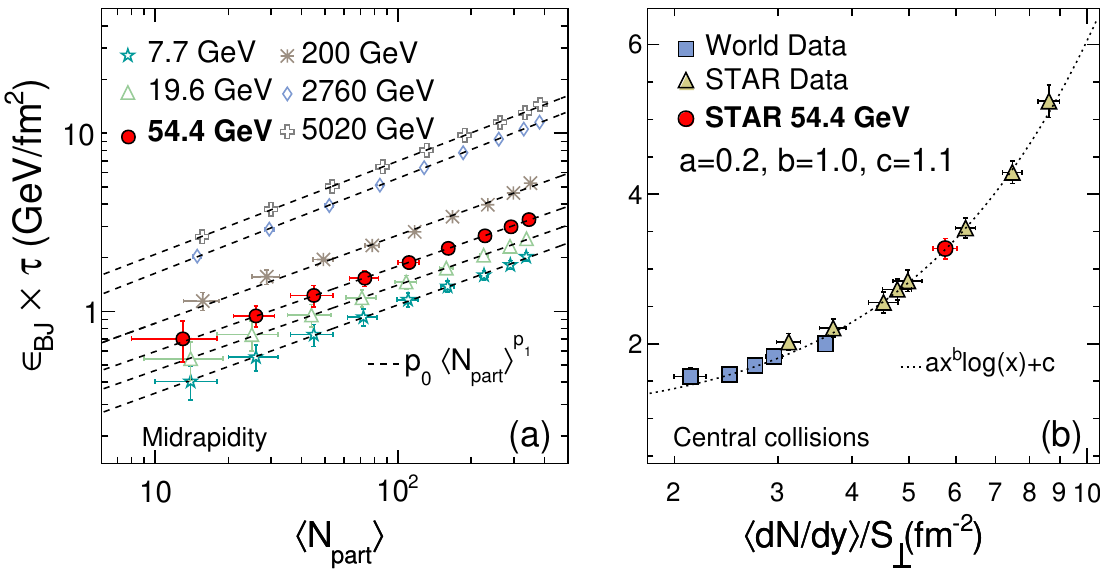}
\vspace{-0.5cm}
\caption{(a) The dependence of $\rm{\epsilon_{BJ}}$ $\times$ $\tau$ as a function of $\langle \rm{N_{part}}\rangle$ for Au+Au collisions at $\sqrt{s_\mathrm{NN}}$ = 54.4~GeV in comparison to the other STAR (Au+Au collisions)~\cite{BES,STAR:2008med} and LHC (Pb+Pb collisions)~\cite{alice2p76,alice5p02} energies. The dashed lines represent the power-law fit at various energies. (b) The variation of $\rm{\epsilon_{BJ}}$ $\times$ $\tau$ with $\frac{\langle dN/dy\rangle}{S_{\perp}}$ for the central collisions at various energies. The dotted line represents the fit to the data.}
\label{ebj}
\end{center}
\end{figure*}

The Bjorken energy density provides insight to the behavior of particle interactions in high-energy collision experiments. It is defined as the energy per unit transverse area in the transverse plane of a high-energy collision. Mathematically,

\begin{equation} \label{eq1}
\rm{\epsilon{_{BJ}}}=\frac{dE_{T}}{dy}\times\frac{1}{S_{\perp}\tau},
\end{equation}
where,
\begin{equation} \label{eq2}
\frac{d E_{T}}{dy} \approx \frac{3}{2}\left(\langle m_{T} \rangle \frac{dN}{dy}\right)_{\pi ^{\pm }} + 2\left(\langle m_{T} \rangle \frac{dN}{dy}\right)_{K ^{\pm }, p, \bar{p}}
\end{equation}
and $S_{\perp}$ is the transverse overlap area of the two colliding nuclei and $\tau$ is the formation time~\cite{STAR:2008med}. The values for $S_{\perp}$ are calculated as- 
\begin{equation} \label{eq3}
{S_{\perp}} = \pi R_{0}^{2} \left(\frac{\langle \rm{N_{part}}\rangle}{2}\right)^{2/3},
\end{equation}
where, $R_{0}$ = 1.2368 fm~\cite{momst}. 
Since the strange particle yields at 54.4~GeV are currently unavailable, $dE_{T}/dy$ is calculated using only the pions, kaons, protons and their corresponding antiparticles. Adding the strange particles (lambdas and cascades) may give rise to 7--8\% increase in the $dE_{T}/dy$ values~\cite{strange_62,strange_bes}. Using~\cref{eq1,eq2,eq3}, the $\epsilon_{\mathrm{BJ}}$ $\times$ $\tau$ is estimated for the STAR~\cite{BES,STAR:2008med} and LHC~\cite{alice2p76,alice5p02} energies and are plotted as a function of $\langle \rm{N_{part}}\rangle$ as shown in Fig.~\ref{ebj}(a). The dashed lines represent the power-law fits of the form $p_{0} \rm{\langle \rm{N_{\text{part}}} \rangle}$$^{p_{1} }$. It is interesting to note that the fit parameter, ${p_1}$, is similar for all energies (see Table~\ref{tab_ebj}). This shows that $\rm{\epsilon_{BJ}}$ $\times$ $\tau$ exhibits same dependence on $\langle \rm{N_{part}}\rangle$ of the system across vast energy range from 7.7 to 5020~GeV which covers a $\mu_{B}$ range from 0 to 400~MeV~\cite{BES}. This suggests that the initial energy density at formation time $\tau$ is similarly distributed across $\langle \rm N_{part} \rangle$ for all the energies. Figure~\ref{ebj} (b) shows the variation of $\rm{\epsilon{_{BJ}}}$ $\times$ $\tau$ as a function of $\langle\frac{dN}{dy}\rangle/S_{\perp}$, where the dotted line represents the fit to the data showing a power-log increase with $\langle\frac{dN}{dy}\rangle/S_{\perp}$. Here, $\langle\frac{dN}{dy}\rangle$ is calculated using
\begin{equation}
\bigg\langle\frac{dN}{dy}\bigg\rangle \approx \frac{3}{2}\left(\frac{dN}{dy}\right)_{\pi ^{\pm }} + 2\left(\frac{dN}{dy}\right)_{K ^{\pm }, p, \bar{p}}.
\end{equation}
For higher energies the increase of $\rm{\epsilon_{BJ}} \times \tau$ as a function of $\langle\frac{dN}{dy}\rangle/S_{\perp}$ is steeper than the lower energies. The lattice QCD predicts the QGP-hadron gas phase transition at energy density 1~GeV/fm$^3$~\cite{lqcda}.The predicted values of the formation time $\tau$ vary across different studies~\cite{tau1, tau2, tau3, STAR:2008med}. For all collision energies considered here, the estimated Bjorken energy density exceeds the phase transition value predicted by lattice QCD using the formation time taken in above references.

\begin{table}
\begin{center}
\begin{tabular}{| c | c | c |}
	\hline
	Energy (GeV) & $p_0$ (GeV/fm$^2$) & $p_1$\\
	\hline 
	7.7 &  0.11 (0.03) & 0.50 (0.04)\\
	\hline
	19.6 & 0.15 (0.04) & 0.48 (0.05)\\
	\hline
	54.4 & 0.20 (0.04) & 0.48 (0.04)\\
	\hline
	200 &0.27 (0.04) &0.49 (0.03)\\
	\hline
	2760 & 0.48 (0.03) &0.53 (0.01)\\
	\hline
	5020 & 0.62 (0.04) &0.52 (0.01)\\
	\hline
\end{tabular}
\caption{The values of the parameters $p_{0}$ and $p_{1}$ obtained by the power-law fitting of Fig.~\ref{ebj}-(a) for various energies is tabulated.} 
\label{tab_ebj}
\end{center}
\end{table}

\begin{figure*}[htbp]
\begin{center}
\includegraphics[scale=0.9]{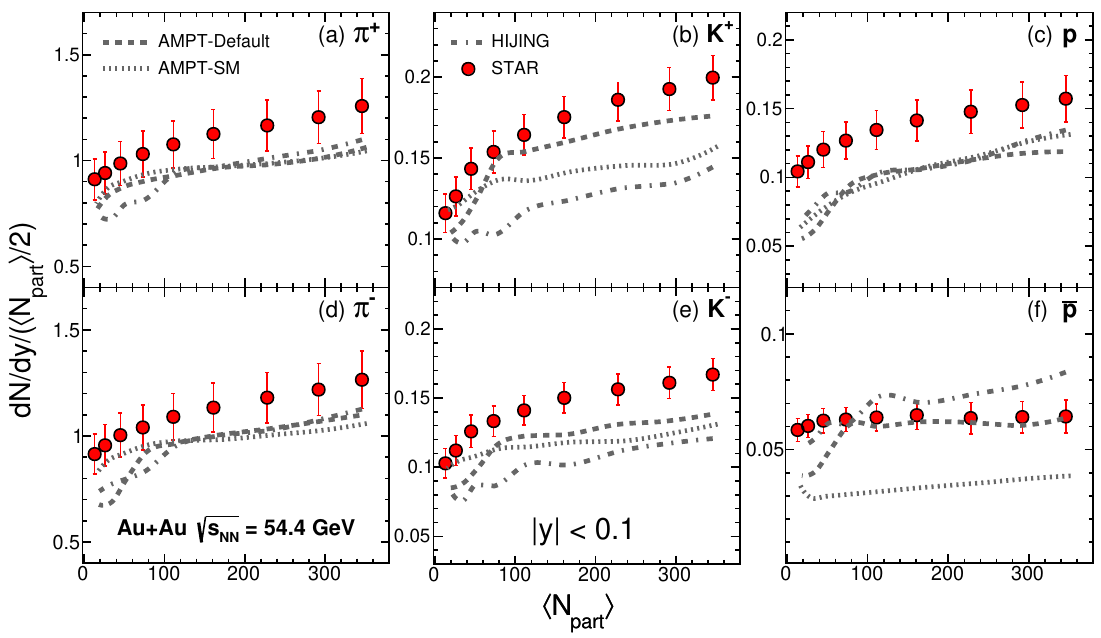}
\vspace{-0.5cm}
\caption{The $\langle \rm{N_{part}}\rangle$ dependence of the normalized integrated particle yield ($dN/dy/(\langle \rm{N_{part}}\rangle/2$) for (a) $\pi^{+}$, (b) $K^{+}$, (c) $p$, (d) $\pi^{-}$, (e) $K^{-}$, (f) $\bar{p}$ at midrapidity ($ |y| < 0.1$) in Au+Au collisions at $\sqrt{s_\mathrm{NN}}$ = 54.4~GeV. The results are compared with AMPT-Default, AMPT-SM and HIJING models.}
\label{defaultdndy}
\end{center}
\end{figure*}

\begin{figure*}[htbp]
\begin{center}
\includegraphics[scale=0.9]{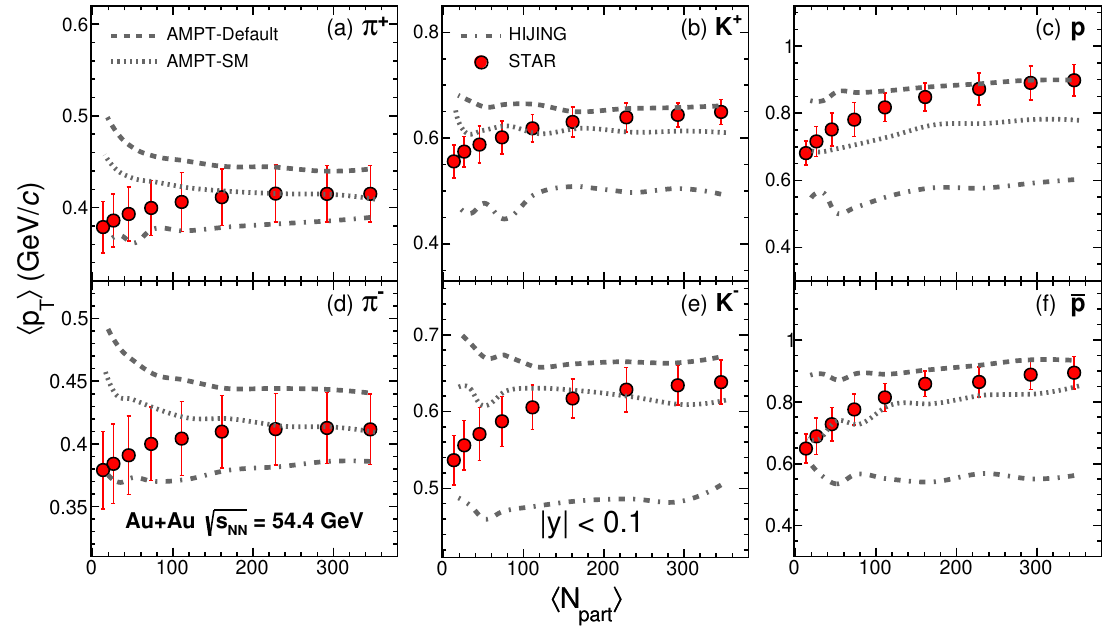}
\vspace{-0.5cm}
\caption{The $\langle \rm{N_{part}}\rangle$ dependence of mean transverse momentum ($\langle p_{T} \rangle$) of (a) $\pi^{+}$, (b) $K^{+}$, (c) $p$, (d) $\pi^{-}$, (e) $K^{-}$, (f) $\bar{p}$ at midrapidity ($ |y| < 0.1$) in Au+Au collisions at $\sqrt{s_\mathrm{NN}} = 54.4$~GeV. The results are compared with AMPT-Default, AMPT-SM and HIJING models.}
\label{meanptmodel}
\end{center}
\end{figure*}

\begin{figure}[htbp]
\begin{center}
\includegraphics[scale=0.41]{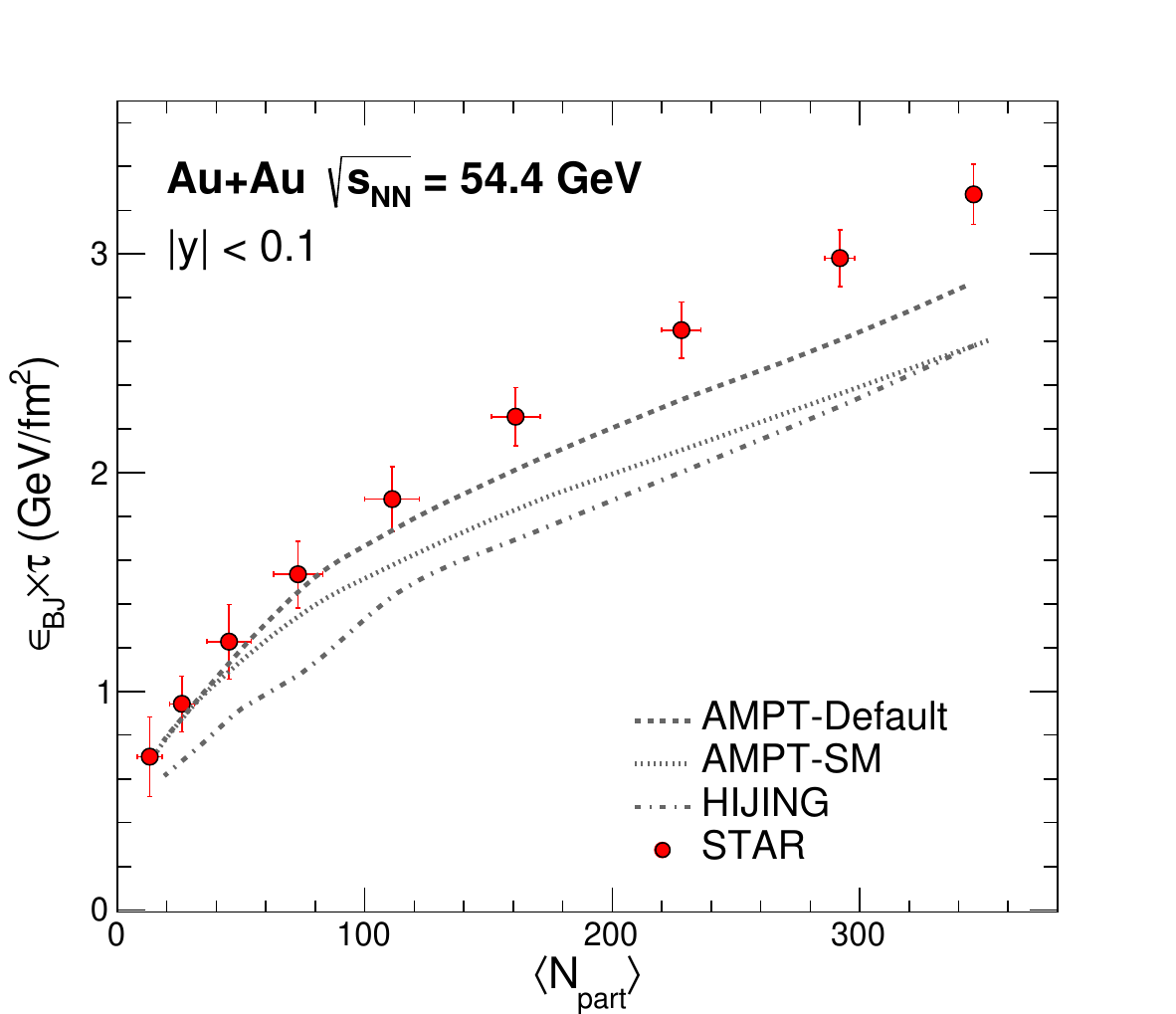}
\vspace{-0.2cm}
\caption{The $\langle \rm{N_{part}}\rangle$ dependence of estimate of $\rm{\epsilon_{BJ}}$ $\times$ $\tau$ at midrapidity ($ |y| < 0.1$) in Au+Au collisions at $\sqrt{s_\mathrm{NN}} = 54.4$~GeV. The results are compared with AMPT-Default, AMPT-SM and HIJING models.}
\label{ebjmodel}
\end{center}
\end{figure}

\section{MODEL COMPARISON}

A Multi-Phase Transport Model (AMPT) simulates the complex interactions that occur in heavy-ion collisions. The initial conditions, including the spatial and momentum distributions of minijet partons and soft excited strings, are generated using the HIJING model. Partonic scatterings are then described by Zhang’s parton cascade (ZPC) model until parton freezeout. In the default version (AMPT-Default), the partons recombine to their parent strings and the resulting strings are converted to hadrons using the Lund string fragmentation model. However, in the string melting version (AMPT-SM), the partons combine into hadrons through a quark coalescence mechanism. Subsequent hadronic scatterings are modeled by A Relativistic Transport (ART) model. Since AMPT includes both initial partonic and final state hadronic interactions and also the transition between these two phases of matter, it is an appropriate model to study various observables in heavy-ion collisions~\cite{ampt,ampt2,ampt3,ampt4}. The Heavy Ion Jet INteraction Generator (HIJING) is a Monte Carlo model that is mainly designed to explore the range of possible initial conditions that may occur in relativistic heavy-ion collisions~\cite{hijing}. For comparison of the models with the experimental data, 50K events were generated for each version of AMPT (version 2.26t9b), and for HIJING (version 1.411). 

Figure~\ref{defaultdndy} shows the variation of integrated yield for each particle as a function of $\langle \rm{N_{part}}\rangle$ compared with those obtained from the AMPT and HIJING models. All the models underestimate the invariant yield of $\pi^{\pm}$ and $K^{\pm}$ for most centralities, and of $p$ for all the centralities. AMPT-Default describes the $\bar{p}$ yield from central to mid central collisions. The results for the $\bar{p}$ yield are overestimated by HIJING for most centralities while AMPT-SM underestimates the same for all centralities.

Figure~\ref{meanptmodel} shows the mean transverse momentum for each particle as a function of $\langle \rm{N_{part}}\rangle$ in comparison with the results obtained for the models. The results from AMPT-SM provide a qualitative description of the mean transverse momentum for $\pi^{\pm}, K^{-}$ from central to mid central collisions, and $\bar{p}$ for almost all the centralities. The values for the $\langle p_T \rangle$ for $K^{+}$ and $p$ are best estimated by AMPT-Default from central to mid central collisions, but are overestimated in peripheral collisions. These data can provide valuable constraints on the AMPT model, particularly in explaining the trend of mean transverse momentum in peripheral collisions.

Figure~\ref{ebjmodel} shows the variation of the estimate of $\rm{\epsilon_{BJ}}$ $\times$ $\tau$ as a function of $\langle \rm{N_{part}}\rangle$ in comparison to the model estimates. AMPT-Default and AMPT-SM models describe the $\rm{\epsilon_{BJ}}$ $\times$ $\tau$ values for a few centralities from peripheral to mid central collisions. The overall description by the AMPT-Default model is better than AMPT-SM. HIJING underestimate the values of $\rm{\epsilon_{BJ}}$ $\times$ $\tau$ as a function of $\langle \rm{N_{part}}\rangle$ for all the centralities.

\section{SUMMARY AND CONCLUSIONS}

We have presented the identified charged particle production in Au+Au collisions at $\sqrt{s_\mathrm{NN}}$ = 54.4~GeV. The transverse momentum spectra of $\pi^{+}$, $\pi^{-}$, $K^{+}$, $K^{-}$, $p$, and $\bar{p}$ in nine centrality classes (0--5\%, 5--10\%, 10--20\%, 20--30\%, 30--40\%, 40--50\%, 50--60\%, 60--70\% and 70--80\%) at midrapidity ($|y|~<~$0.1) were obtained. The centrality and energy dependence of particle yield,  $\langle p_{T}\rangle$ or $\langle m_{T}\rangle$, and particle yield ratios were studied. The kinetic freezeout parameters that govern the system dynamics were extracted and Bjorken energy density times the formation time ($\rm{\epsilon_{BJ}} \times \tau$), which represents the energy density in the central rapidity region of the collision zone, was estimated. 

The transverse momentum spectra flattens for the higher mass particles. In addition, $\langle p_{T}\rangle$ increases towards the central collisions, suggesting the presence of radial flow in these collisions. $dN/dy/(\langle \rm N_{part} \rangle/2)$ as a function of $\langle \rm N_{part} \rangle$ increases from peripheral to central collisions for $\pi^{\pm}$, $K^{\pm}$ and $p$. This suggest that there are some contributions from hard processes that involve nucleon-nucleon binary collisions. However, $\bar{p}$ shows a weak centrality dependence. 

In Au+Au collisions at $\sqrt{s_\mathrm{NN}} = 54.4$~GeV, the $\pi^{-}/\pi^{+}$ ratio remains close to unity across all centralities. The $K^{-}$/$K^{+}$ ratio is almost flat as a function of $\langle \rm N_{part} \rangle$, suggesting that $K^{+}$ and $K^{-}$ production have similar centrality dependence. The $K^{\pm}/\pi^{\pm}$ ratios rise from peripheral to mid central collisions and then saturate towards central collisions. The centrality dependence of $K^{+}/\pi^{+}$ ratio is steeper at lower energies, compared to those at higher energies, while that for $K^{-}/\pi^{-}$ shows similar dependence at all energies.

The $\bar{p}/p$ ratio at 54.4~GeV shows a slight decrease towards the central collisions. This decreasing behavior is more prominent at lower STAR energy 7.7~GeV. The $p/\pi^{+}$ ratio at 54.4~GeV shows a little rising trend from peripheral to central collisions. This increase is more steep for lower energy 7.7~GeV. The $\bar{p}/\pi^{-}$ ratio shows slight decrease from peripheral to central collisions. All these observations can be attributed to baryon stopping at midrapidity and/or baryon-antibaryon annihilation in more central collisions.

The energy dependence of particle yields in the most central Au+Au collisions at 54.4~GeV in midrapidity follows the trend of other energies. At 54.4 GeV, the ratio $\pi^{-}/\pi^{+}$ $\sim$ 1, indicating that the main production mechanism for these particles is pair production. The $K^{-}/K^{+}$ $\sim$ 0.84 suggesting that there is a contribution from associated production at 54.4~GeV and $\bar{p}/p$ $\sim$ 0.40 reflecting baryon stopping in these collisions at midrapidity. At 54.4~GeV, the $K^{\pm}/\pi^{\pm}$ as a function of energy follows the trend of other energies. The $K^{+}/\pi^{+}$ ratio exhibits a horn like structure as a function of energy at around 7.7~GeV. Moreover, it is observed that the $K^{-}/K^{+}$ and $\bar{p}/p$ ratios are correlated and can be described by a power-law function with an exponent of approximately 0.2.

The $\langle m_{T}\rangle - m$ for different particles increases with $\sqrt{s_\mathrm{NN}}$ at lower energies, and then becomes constant around the BES energies (7.7--39~GeV) and rises further for higher energies. The flat behavior has been attributed to the existence of mixed phase of QGP and hadrons~\cite{BES}. At 54.4~GeV, its value is above this constant trend which may reflect that the partonic degrees of freedom might be dominating at this energy.

The kinetic freezeout parameters were obtained using the hydrodynamics based blast-wave model. The extracted $T_\text{kin}$ and $\langle \beta_{T} \rangle$ parameters show an anticorrelation with each other. The increasing $\langle \beta_{T} \rangle$ and decreasing $T_\text{kin}$ values towards central collisions could be an indication of more rapid expansion of the system and a longer-lived fireball in such collisions. 

The $\rm{\epsilon_{BJ}} \times \tau$, was estimated and plotted as a function of number of participating nucleons in various collision energies. It is noted that the dependence of $\rm{\epsilon_{BJ}} \times \tau$ on $\langle \rm N_{part} \rangle$ is same for all energies ranging from lower STAR energy of 7.7~GeV to higher LHC energy of 5020~GeV, suggesting that the initial energy density at formation time $\tau$ depends on $\langle \rm N_{part} \rangle$ in the same manner across various energies. At $\sqrt{s_{\rm{NN}}}=54.4$~GeV, the Bjorken energy density exceeds the value predicted by lattice QCD for phase transition.

The results obtained were compared with the model estimates of AMPT-Default, AMPT-SM and HIJING. Comparisons with these models showed varying degrees of agreement with the data, highlighting the complexities of modeling particle production in heavy-ion collisions.
\vspace{0.5cm}

\section{ACKNOWLEDGEMENTS}
We thank the RHIC Operations Group and SCDF at BNL, the NERSC Center at LBNL, and the Open Science Grid consortium for providing resources and support.  This work was supported in part by the Office of Nuclear Physics within the U.S. DOE Office of Science, the U.S. National Science Foundation, National Natural Science Foundation of China, Chinese Academy of Science, the Ministry of Science and Technology of China and the Chinese Ministry of Education, NSTC Taipei, the National Research Foundation of Korea, Czech Science Foundation and Ministry of Education, Youth and Sports of the Czech Republic, Hungarian National Research, Development and Innovation Office, New National Excellency Programme of the Hungarian Ministry of Human Capacities, Department of Atomic Energy and Department of Science and Technology of the Government of India, the National Science Centre and WUT ID-UB of Poland, German Bundesministerium f\"ur Bildung, Wissenschaft, Forschung and Technologie (BMBF), Helmholtz Association, Ministry of Education, Culture, Sports, Science, and Technology (MEXT), Japan Society for the Promotion of Science (JSPS), and Agencia Nacional de Investigacion y Desarrollo de Chile (ANID), Chile.

\bibliography{reference2}
\end{document}